\documentclass[11pt]{IEEEtran}
\usepackage{graphicx,xcolor,cite} 
\usepackage{romannum,wrapfig,float,amsmath,amsfonts}
\usepackage[bookmarks,colorlinks]{hyperref}
\usepackage[subtle]{savetrees}
\begin{document}

\title{In-Context Learning for Gradient-Free Receiver Adaptation: Principles, Applications, and Theory}

\author{Matteo Zecchin, Tomer Raviv, Dileep Kalathil, Krishna Narayanan, Nir Shlezinger, and Osvaldo Simeone
\thanks{Matteo Zecchin and Osvaldo Simeone are with the King’s Communications, Learning and Information Processing (KCLIP) lab within  the Centre for Intelligent Information Processing Systems (CIIPS) at the Department of Engineering,  King’s College London, London, WC2R 2LS, UK. (email: \{osvaldo.simeone, matteo.1.zecchin\}@kcl.ac.uk). Tomer Raviv and Nir Shlezinger are with the School of Electrical and Computer Engineering at Ben-Gurion University
	 (email: \{tomerraviv95, nirshlezinger1\}@ece.tamu.edu).
	Dileep Kalathil and Krishna Narayanan are with the Department of Electrical and Computer Engineering, Texas A\&M University (email: \{dileep.kalathil, krn\}@ece.tamu.edu). The work of Matteo Zecchin and Osvaldo Simeone was supported by the European Union’s Horizon Europe project CENTRIC (101096379),  by~an Open Fellowship of the EPSRC (EP/W024101/1), and~by the EPSRC project (EP/X011852/1). The work of 	Dileep Kalathil and Krishna Narayanan is supported in part by the National Science Foundation under grant 2148354 and in part by funds from federal agency and industry partners as specified in the Resilient \& Intelligent NextG Systems (RINGS) program.}
}

\maketitle
\vspace{-4em}
\begin{abstract}
   In recent years, deep learning has facilitated the creation of wireless receivers capable of functioning effectively in conditions that challenge traditional model-based designs. Leveraging programmable hardware architectures, deep learning-based receivers offer the potential to dynamically adapt to varying channel environments. However, current adaptation strategies, including joint training, hypernetwork-based methods, and meta-learning, either demonstrate limited flexibility or necessitate explicit optimization through gradient descent.

This paper presents gradient-free adaptation techniques rooted in the emerging paradigm of in-context learning (ICL). We review architectural frameworks for ICL based on Transformer models and structured state-space models (SSMs), alongside theoretical insights into how sequence models effectively learn adaptation from contextual information. Further, we explore the application of ICL to cell-free massive MIMO networks, providing both theoretical analyses and empirical evidence. Our findings indicate that ICL represents a principled and efficient approach to real-time receiver adaptation using pilot signals and auxiliary contextual information—without requiring online retraining. 
\end{abstract}
\section{Context and  Motivation}

\begin{figure*}
    \label{fig:taxonomy_neural_receivers}
    \centering
    \includegraphics[width=0.9\linewidth]{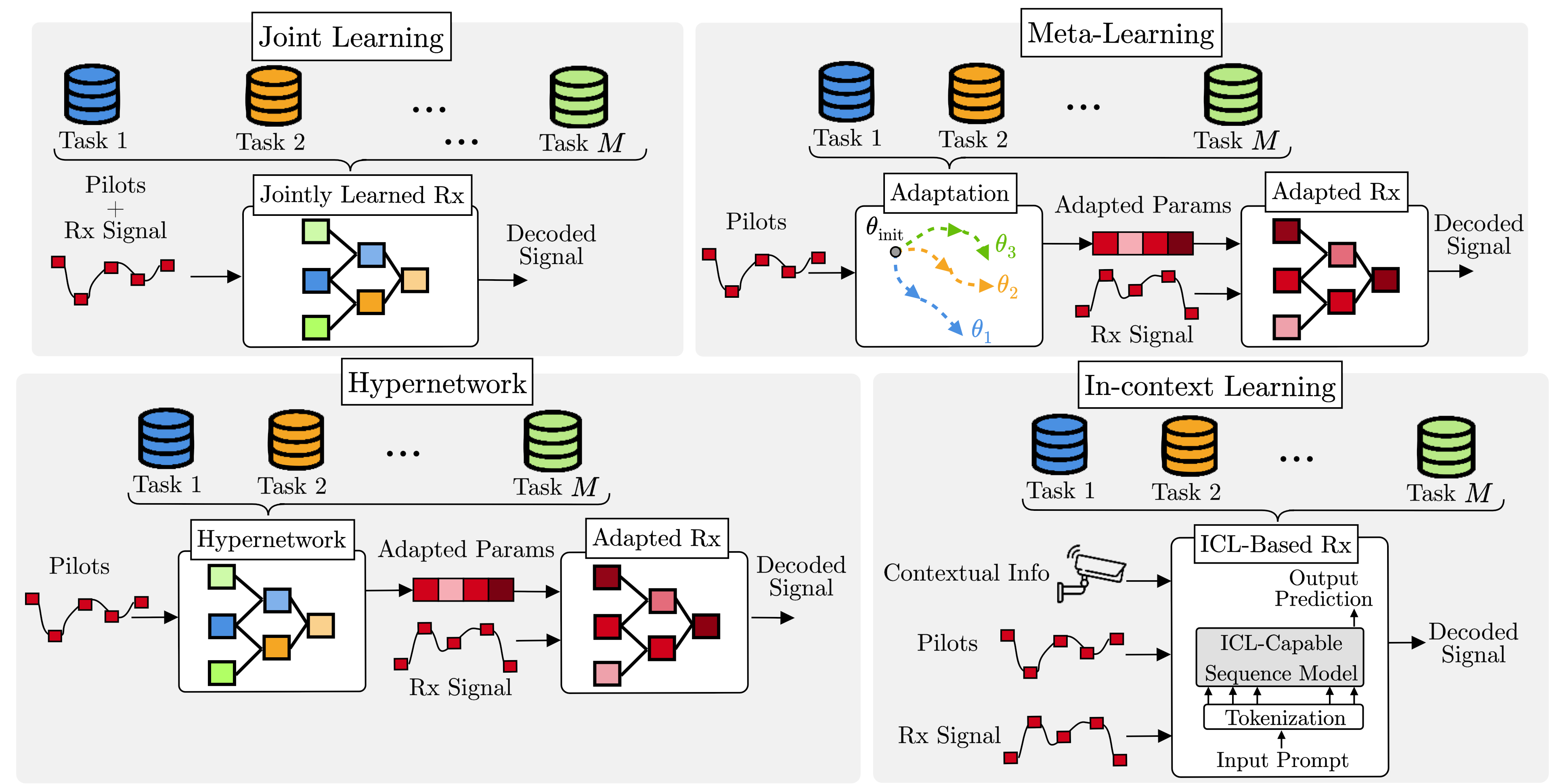}
    \caption{The challenge of ensuring the adaptivity of neural receivers to varying traffic, channel, and environmental conditions has been addressed using different design principles: joint learning, meta-learning, hypernetworks, and in-context learning (ICL). ICL has unique potential benefits, as it does not require explicit optimization at run time, and it can seamlessly incorporate complex contextual information such as network topology, large-scale fading statistics, and interference characteristics.}
    \label{fig:neural_receivers_overview}
\end{figure*}

\subsection{The Need for Adaptivity}
Traditionally, wireless receivers have been designed by relying on \emph{mathematical models} for the cascade of transmitter circuitry, propagation channel, and receiver front-end. Conventional models account for aspects such as I/Q imbalance, time and frequency selectivity due to multipath channels,  white Gaussian noise, and receiver non-linearities. \emph{Adaptation} to changing environments is typically achieved through the transmission of known signals (e.g., pilots), which support the estimation of end-to-end channel responses. In modern network settings, this traditional approach is challenged by hard-to-model factors such as interference from co-channel services \cite{testolina2024modeling} and pilot contamination \cite{bjornson2019making},  as well as by the complexity of optimized solutions for signal models involving new features such as intelligent reflecting surfaces \cite{karasik2021adaptive}, and reuse of radio signals for sensing \cite{wymeersch2024joint}.

To address the outlined limitations of model-based solutions, \emph{data-driven}  designs using machine learning tools have emerged as viable solutions \cite{raviv2024adaptive}. Notably,  recent advances have demonstrated the feasibility of competitive standard-compliant implementations of neural receivers \cite{wiesmayr2024designstandardcompliantrealtimeneural}.  One of the key technical challenges in designing neural receivers is the need to ensure \emph{adaptivity} to varying traffic, channel, and environmental conditions by incorporating pilots and contextual information, while abiding by constraints on time, data, and computational resources at the physical layer. Contextual information may encompass network topology, large-scale fading statistics,  interference characteristics, or sensory data, such as video streams from cameras.

\subsection{From Joint Learning to In-Context Learning}

As illustrated in Figure~\ref{fig:neural_receivers_overview}, several techniques have been proposed to enable adaptivity in data-driven approaches (see also \cite{simeone2022machine} for an overview). A simplistic approach would be to apply a continual, or online, learning methodology that requires re-training at each step, typically with some anchoring to previously trained models to avoid issues such as catastrophic forgetting. 

More practical and scalable solutions that do not require the re-training of the base model rely on a pre-training step that leverages the availability of data from multiple  \emph{tasks}, where each task represents specific traffic patterns, channel characteristics, and environmental conditions. Although the specific design implementations vary, all these methods fundamentally aim to optimize adaptability across a diverse set of tasks:
\begin{itemize}
     \item \textit{Joint learning} produces a one-size-fits-all neural receiver by pooling together the available data sets for all tasks. Adaptation to a new task is based on processing estimated channels as part of the input of the model. Due to the lack of specialization of the inner working of the neural receiver to current conditions, a jointly learned receiver can perform poorly, being unable to extract sufficient information from the pilot signals (see, e.g., \cite{chen2023learning}).
     \item \textit{Meta-learning}, or \emph{learning to learn}, is a training approach that targets neural receivers explicitly designed to enable fast adaptation and potentially facilitate online learning~\cite{raviv2023online}. Still, meta-learning relies on complex optimization routines, typically based on gradient descent, for each  received pilot sequence, limiting their practical robustness and performance in wireless systems (see, e.g., \cite{zecchin2024cell}).
     \item \textit{Hypernetworks} are neural networks that are trained to output the parameters of another neural network \cite{ha2016hypernetworks}. Neural receiver adaptation is achieved by feeding a pilot sequence into a  hypernetwork, which returns the adapted parameters for the neural receiver. The received signal is then decoded using the adapted neural receiver \cite{raviv2024modular}. Unlike meta-learning, hypernetworks do not require any run-time optimization. However, due to the complex relationship between a neural receiver’s parameters and its performance, hypernetworks are often difficult to train, and they tend to provide adaptivity to relatively limited variations. Furthermore, it is generally hard to flexibly incorporate additional contextual information as the input of hypernetworks.

\end{itemize}

In this article, we review a new design principle that leverages the power of modern sequence models to carry out \emph{in-context learning} (ICL) \cite{xie2021explanation}.

\subsection{In-Context Learning}
ICL refers to a model’s ability to adapt to a new task by processing a sequence of examples provided as part of its input, also known as \emph{prompt}. Accordingly, as shown in Figure~\ref{fig:neural_receivers_overview}, when applied to the design of neural receivers, ICL implements a direct mapping from a prompt consisting of pilot symbols, contextual information, and current received signals to the decoded output. 

Conceptually, ICL shares with meta-learning an operation characterized by a separation of time scales: a \emph{slow learning} process that leverages data from multiple tasks to support a \emph{fast learning} process based on context at run time. However, unlike meta-learning, adaptation -- fast learning -- does not depend on fragile explicit optimization routines, relying instead on a direct mapping.  Accordingly, at run time, the ICL-based receiver adapts in a \emph{gradient-free} manner, without requiring the  fine-tuning of the receiver's parameters.  ICL-based mappings are  typically implemented using Transformer models, though other \emph{sequence models} -- notably \emph{state-space models} (SSMs) also exhibit ICL capabilities  \cite{park2024can}.

\begin{figure*}
    \centering
    \includegraphics[width=0.95\linewidth]{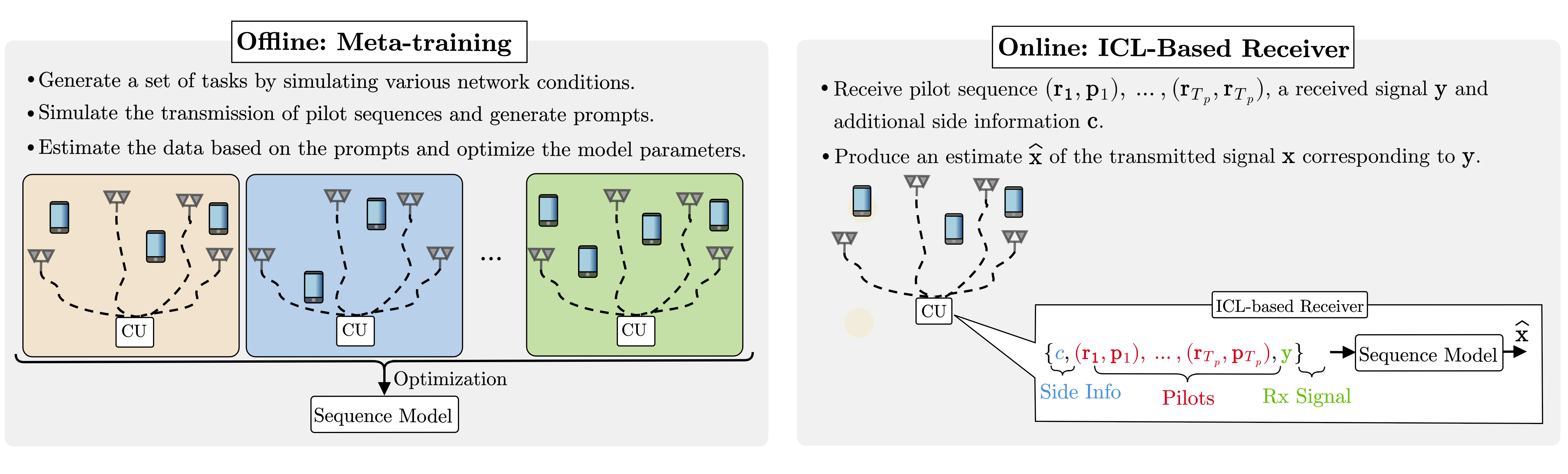}
    \caption{An ICL-based receiver is optimized during an offline meta-training phase that leverages data from multiple tasks, i.e., multiple traffic, channel, and environmental conditions. At run time, an ICL-based receiver adapts in a gradient-free manner, without requiring the fine-tuning of the receiver's parameters, using solely pilots, contextual information, and the current received signal to produce the estimated transmitted signal.}
    \label{fig:icl}
\end{figure*}

As summarized in Figure~\ref{fig:icl}, ICL capabilities are developed during a pre-training phase, in which data -- contextual information, along with transmitted and received pilot signals -- from multiple tasks are leveraged to optimize the mentioned mapping from prompt to decoded signal. 

\subsection{This Paper}

The main features of this paper are as follows:
\begin{itemize}
\item \emph{ICL for gradient-free adaptation}: We introduce and review ICL-based wireless receivers. Against the backdrop of existing design strategies for adaptive receivers, including joint learning, hypernetwork-based adaptation and  meta-learning, we present ICL as an emerging paradigm in which the model itself learns to adapt internally from examples provided in its input prompt.  ICL enables  sequence model to effectively infer the underlying channel or task from a few pilot symbols and then perform the required signal processing task without any explicit weight updates.
\item \emph{Model architectures for ICL}:  We describe how architectures like Transformers and modern SSMs can be employed to enable ICL, and  discuss training methodologies that imbue these models with in-context adaptivity. 
\item \emph{Theoretical insights into ICL}: We review theoretical insights into ICL in terms of optimal Bayesian inference,  providing a principled explanation for how ICL emerges from appropriate meta-training (see Figure~\ref{fig:icl}).
\item \emph{Applications to MIMO systems}: We examine application scenarios in cell-free massive MIMO networks to illustrate the performance of ICL and highlight how ICL-based receivers leverage pilot signals and other contextual information to achieve real-time adaptation without any parameter fine-tuning.

\end{itemize}

The remainder of this paper is organized as follows. Section~\ref{sec:icl_general} provides a general description of the ICL framework, along with an overview of existing architecture models that are ICL-capable and their associated training techniques, contrasting them with traditional methods. 

Section \ref{sec:prob_formulation} introduces the cell-free massive MIMO setting that will be used to demonstrate the application of ICL to the problem of channel equalization, and reviews state-of-the-art adaptive receivers. Section \ref{sec:icl_receiver} presents  ICL-based neural receivers, detailing their implementation, training procedure, and theoretical properties. Section \ref{sec:icl_results} presents a series of experiments investigating: (\emph{i}) the relation between meta-training task diversity and the behavior of ICL-equalizers; (\emph{ii}) the computational and memory requirements of Transformer-based and SSM-based ICL-equalizers; (\emph{iii}) the performance of the ICL-equalizer relative to existing adaptive receivers; and (\emph{iv}) the out-of-distribution robustness of ICL receivers. Section \ref{sec:outlook} discusses current and future research directions in the design, application, and analysis of ICL-based neural receivers. Finally, Section \ref{sec:conclusions} concludes the paper.

\section{In-Context Learning}
\label{sec:icl_general}
ICL refers to a model’s ability to adapt to a new task by processing a sequence of examples provided as part of its input, without updating its parameters~\cite{xie2021explanation}.  The core principle is that a sufficiently large and well-trained model can internalize a learning algorithm in its forward pass.

\subsection{General ICL Framework}
\subsubsection{Tasks and Contexts} Assume a family of tasks $\mathcal{T}$, each corresponding to a different data-generating process. Different tasks may correspond, for example, to distinct propagation environments, communication parameters, and/or the number of active users in the system. In ICL, a model is trained on many tasks from family $\mathcal{T}$ such that, at inference time, it can quickly adapt to a new task from the same family  $\mathcal{T}$.

Adaptation to a new task is based on contextual information $\mathcal{C}$. For a task $\tau \in \mathcal{T}$, this typically consists of a set of $N$ i.i.d. input-output examples ${(u_i, v_i)}_{i=1}^N$ from the task data-generating distribution $P_{u,v|\tau}$. The context may also include additional information describing the task -- such as a textual characterization of the propagation environment.

For a task $\tau$, given context $\mathcal{C}$ and a new input $u\sim P_{u|\tau}$, an ICL-capable model produces the estimate $\hat{v}$ of the corresponding output $v\sim P_{v|u,\tau}$. Crucially, the model’s weights remain fixed during this process, and any adaptation to task $\tau$ happens implicitly within the forward pass. This way, rather than being explicitly encoded within an optimization routine, adaptation is learned by the sequence model during a pre-training phase, effectively making the model itself an optimizer. For example, recent theoretical work has shown that Transformers can implement linear regression in-context through forward-only adaptation, without the need to carry out gradient descent \cite{garg2022can}. 

\subsubsection{Transformer Architectures for ICL} 
\begin{figure}
    \centering
    \includegraphics[width=0.7\linewidth]{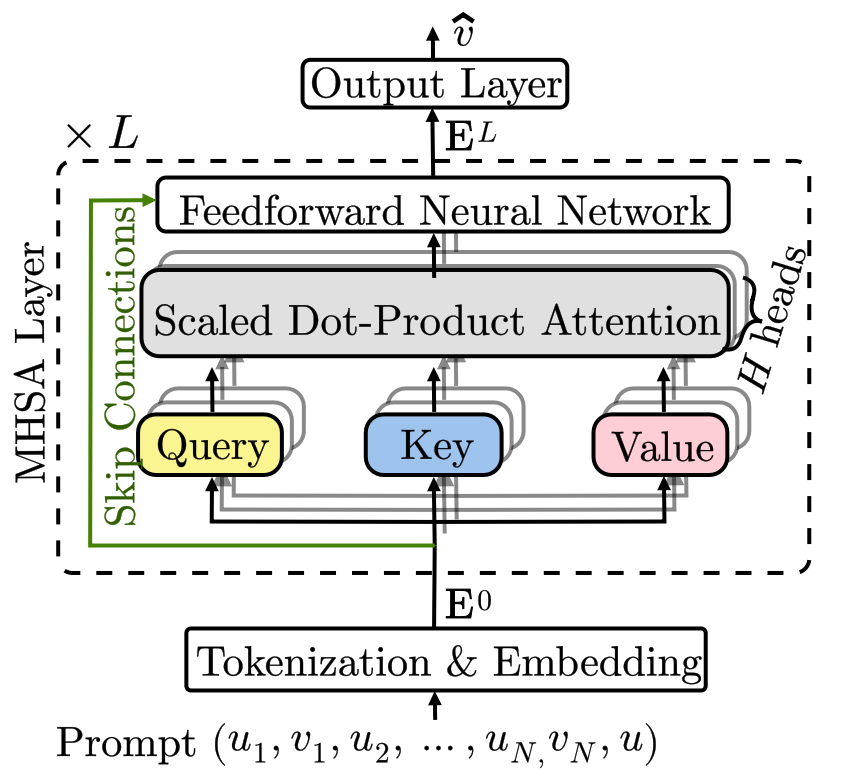}
    \caption{A Transformer obtains the estimated output $\hat{v}$ by first tokenizing the in-context examples $(u_1, v_1, \dots, u_N, v_N)$ and the query input $u$, and then processing the resulting embeddings through repeated application of the MHSA layer, followed by an output neural layer. }
    \label{fig:tf}
\end{figure}

While any sequence model could in principle be trained to exhibit ICL, decoder-only Transformer-based sequence models have been the first to show a remarkable capacity for it. Transformers are well-suited for ICL, as they can process a concatenated sequence of context examples, attending to all context items to inform the prediction on the query. Specifically, as illustrated in Figure~\ref{fig:tf}, a Transformer produces the prediction $\hat{v}$ by processing the input sequence, also known as \emph{prompt},
\begin{equation}
(\mathcal{C}
,u)=(u_1,v_1,u_2,v_2,…,u_N,v_N,u),
\label{eq:input_seq}
\end{equation}
via the repeated applications of a multi-head self-attention (MHSA) layer \cite{turner2023introduction}.

\begin{figure}
    \centering
    \includegraphics[width=0.65\linewidth]{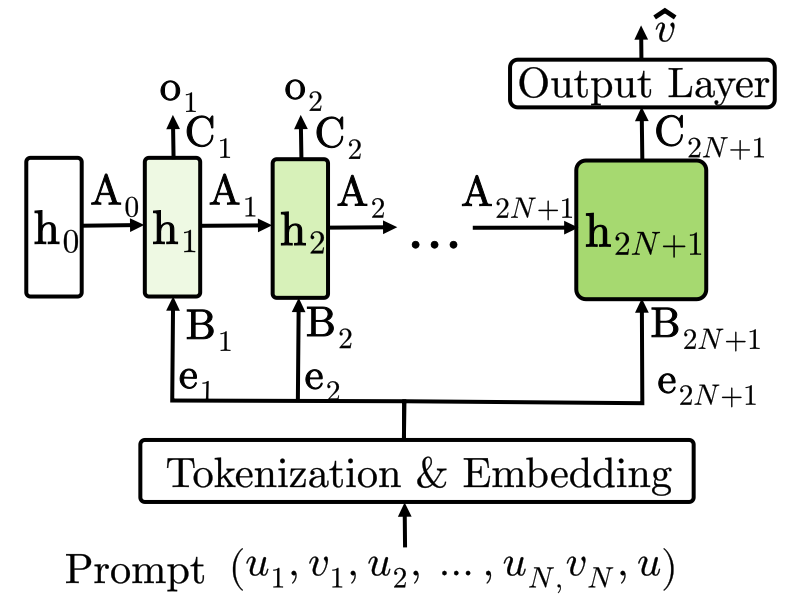}
    \caption{A State Space Model operates sequentially on the tokenized sequence of in-context examples $(u_1,v_1,\dots,u_N,v_N)$ and query input $u$. Each token is used to perform sequential linear updates to a hidden state $\mathbf{h}_{t}$ at each step. The final prediction $v$ is derived based on the value hidden state after the entire sequence has been processed. }
    \label{fig:ssm}
\end{figure}

The raw input sequence \eqref{eq:input_seq} includes input and output examples. The input examples are typically real-valued vectors, e.g., received signal samples, while output examples may be categorical variables, e.g., decoded signals.  A preliminary step toward applying the MHSA mechanism is \textit{tokenization}, which transforms the input data sequence into a series of real-valued vectors, known as tokens, all with a common dimension $D_t$. Tokenization is usually applied independently to each element of the prompt \eqref{eq:input_seq}, yielding a sequence of $2N+1$ tokens. However, alternative approaches, where input and output examples are jointly tokenized into a single token, are also possible \cite{garg2022can}.  Unlike standard applications of attention, positional encodings is typically not applied in ICL, as the task is invariant to the order of the example pairs in the prompt \cite{fang2025rethinking}.

The tokenized sequence is then organized into a matrix $\mathbf{T} \in \mathbb{R}^{D_t \times (2N+1)}$, where each column corresponds to the tokenized vector associated with an element of the sequence \eqref{eq:input_seq}. Each token is projected into an embedding space of dimension $D_e$ via the column-wise application of a trainable linear map to the matrix $\mathbf{T}$. This yields an initial embedding matrix $\mathbf{E}^0 \in \mathbb{R}^{D_e \times (2N+1)}$, which serves as the input to the first MHSA layer. Each subsequent layer $l \geq 1$ then operates on the output $\mathbf{E}^{l-1} \in \mathbb{R}^{D_e \times (2N+1)}$ from the previous layer.

The MHSA mechanism applies $H$ attention heads, with each head $h$ parameterized by key, query, and value matrices. These are denoted as $\mathbf{W}^K_h \in \mathbb{R}^{D_w \times D_e}$, $\mathbf{W}^Q_h \in \mathbb{R}^{D_w \times D_e}$, and $\mathbf{W}^V_h \in \mathbb{R}^{D_v \times D_e}$, respectively, where $D_w$ is the dimension of the key and query representations, while $D_v$ is the dimension of the value representations. Typically, the dimension $D_w$ is set to be smaller than the dimension $D_e$ to reduce the computational cost of the attention mechanism. Each column of the $D_v \times (2N+1)$ matrix $\mathbf{O}_h^l$ produced by the $h$-th attention layer $l$ is a convex combination of the corresponding value representation output by layer $l-1$, weighted by query-key similarity, typically via a scaled dot-product of the form 
\begin{equation}
\label{eq:attention}
\mathbf{O}_h^l = \mathbf{W}^V_h \mathbf{E}^{l-1} \cdot \text{softmax}\left(\frac{(\mathbf{W}^K_h \mathbf{E}^{l-1})^\top (\mathbf{W}^Q_h \mathbf{E}^{l-1})}{\sqrt{D_w}}\right),
\end{equation}
where the softmax function is applied column-wise \cite{turner2023introduction}. The operation \eqref{eq:attention} is also known as \textit{softmax attention}. The outputs of the $H$ attention heads are concatenated and processed by a feed-forward network with residual connections to produce the embedding matrix for the next layer $\mathbf{E}_l$.

After applying the MHSA layer $L$ times, the resulting embedding sequence $\mathbf{E}_L$ is used to produce the estimate $\hat{v}$. This is usually done by extracting the final column of $\mathbf{E}_L$ and passing it through a trainable feedforward neural network.

Due to the need to compute and store the attention matrix \eqref{eq:attention},  Transformers have a computational complexity at inference time that scales \textit{quadratically} with the size of the context. This may pose a problem in settings characterized by complex tasks requiring rich contextual information for adaptation. 

\subsubsection{SSM Architecture for ICL}
A class of sequence models that addresses the quadratic complexity of Transformers, while still exhibiting ICL capabilities \cite{park2024can}, is given by SSMs. At their core, SSMs are linear gated recurrent models. As such, they are strongly related to Transformer models with \textit{linear}, rather than softmax, \textit{attention},  and can be interpreted as a form of SSM with a particular structure of the state matrix.

As shown in Figure~\ref{fig:ssm}, the input sequence \eqref{eq:input_seq} is first tokenized and then projected into an embedding sequence $\mathbf{E} \in \mathbb{R}^{D_e \times (2N+1)}$ as in the case of Transformers. However, unlike Transformers, the embedding matrix $\mathbf{E}$ is not processed jointly. Instead, its columns $(\mathbf{e}_1, \mathbf{e}_2, \dots, \mathbf{e}_{2N+1})$ are used to sequentially update a hidden state $\mathbf{h} \in \mathbb{R}^{D_h}$ with dimension $D_h$. Specifically, each column $\mathbf{e}_t$ is used to update $\mathbf{h}_{t-1}$ according to the discrete-time update dynamics
\begin{equation}
\label{eq:mamba}
\mathbf{h}_t = \mathbf{A}_t \mathbf{h}_{t-1} + \mathbf{B}_t \mathbf{e}_t,
\end{equation}
where the sequence of matrices $\mathbf{A}_t \in \mathbb{R}^{D_h \times D_h}$ is fixed, while the sequence of matrices $\mathbf{B}_t \in \mathbb{R}^{D_h \times D_e}$ is allowed to be input-dependent and is typically produced by a trainable feedforward neural networks applied to the embedding $\mathbf{e}_t$.

At every time step $t$, the hidden state $\mathbf{h}$ is mapped to an output vector $\mathbf{o}_t \in \mathbb{R}^{D_o}$ of dimension $D_o$ via the linear map
\begin{equation}
\label{eq:mamba_output}
\mathbf{o}_t = \mathbf{C}_t \mathbf{h}_t,
\end{equation}
where matrix $ \mathbf{C}_t$ is allowed to depend on $\mathbf{e}_t$ in a manner similar to $\mathbf{B}_t$. The prediction $\hat{v}$ is typically obtained from the last output $\mathbf{o}_{2N+1}$ via a trainable feedforward neural network.

Due to the linear autoregressive nature of the SSM dynamics in \eqref{eq:mamba}, the complexity of SSMs scales linearly with the size of the context. 

\subsubsection{Prompt Design} Prompt design is critical to facilitate adaptation via ICL. In language tasks such as text classification, question answering, and text summarization, the context is typically presented alongside additional information about the task to be solved. This contextual information is provided as additional tokens, often preceding those representing the examples, allowing the model to adapt more effectively to the task at hand. In Section~\ref{sec:icl_receiver}, we illustrate this principle in the context of channel equalization in cell-free massive MIMO systems, where providing contextual information -- such as long-term channel statistics -- can significantly improve adaptation.

 \subsection{ Pre-Training Strategies for ICL} 
 \label{sec:icl_gen_train}
An ICL-capable model is typically obtained via a \emph{meta-training} procedure based on the family of tasks $\mathcal{T}$.  The training procedure consists of the following steps, repeated across multiple iterations: \begin{enumerate}
\item \emph{Task generation}: For each training iteration, a set of tasks is sampled from the task distribution. This step may involve simulating, for example, a communication system under different channel conditions, traffic loads, user activation patterns, and communication parameters. The task distribution should ideally cover the kinds of variations expected at test time.
\item \emph{Prompt Generation}: For each sampled task $\tau$, a number of prompts are generated by sampling input-output examples from distribution $P_{u,v|\tau}$. 
\item \emph{Optimization}: The model processes each prompt \eqref{eq:input_seq} and outputs an estimate $\hat{v}$ of the output $v$ corresponding to the query input $u$. A loss function $\ell(v,\hat{v})$ -- e.g., the mean squared error or the cross-entropy -- is evaluated on the model’s output $\hat{v}$ and the true output $v$. The loss is aggregated over prompts and tasks. The aggregated loss is optimized over the model weights via first order optimization methods, such as stochastic gradient descent. This optimization trains the ICL model to interpret context examples and produce accurate query outputs, effectively teaching it to adapt to the task on the basis of the given examples.

\end{enumerate}

\subsection{Comparison with Related Frameworks}
\label{sec:icl_vs_rest}
\subsubsection{ICL vs. Few-Shot Meta-Learning} The training process reviewed above follows the same basic steps as for meta-learning \cite{chen2023learning}. However, while ICL directly optimizes the forward inference process, meta-learning optimizes fine-tuning based on contextual information. As a result, typical forms of meta-learning -- such as model-agnostic meta-learning (MAML) \cite{finn2017model} -- require the evaluation of second-order derivatives to capture the sensitivity of the fine-tuning process to the parameters being optimized. Furthermore, this also necessitates the specification of optimization hyperparameters such as learning rate, batch size, and number of adaptation steps.

Given this fundamental distinction with meta-learning, ICL is considered to be a form of \emph{mesa-learning}. Mesa-learning is generally said to  occur when a base optimizer  -- here the meta-training process -- trains a model that itself learns a new internal optimization process. So, while meta-learning effectively learns to actively update a model in order to carry out a new task,  mesa-learning automatically adjusts to new contexts without requiring any active learning.

\subsubsection{ICL vs. Joint Learning and Hypernetworks}
As shown in Figure~\ref{fig:neural_receivers_overview}, joint learning trains a single model by pooling data from all sampled tasks, with the aim of optimizing average performance across tasks. At inference time, the trained model is applied directly to new tasks without explicit task-specific adaptation.

In contrast, a hypernetwork is a neural network that generates the weights of another network based on input contextual variables. By using the weights provided by a hypernetwork, the target model can adapt to different tasks or environments, typically without requiring gradient-based updates at test time.

ICL can be seen as combining aspects of both paradigms: like joint learning, it yields a single model applicable to many tasks; and, like hypernetworks, it uses task-specific input -- in the form of example input-output pairs -- to adapt its behavior during inference.

\section{Equalization in Cell-Free Massive MIMO Systems}

\begin{figure}
    \centering
    \includegraphics[width=0.7\linewidth]{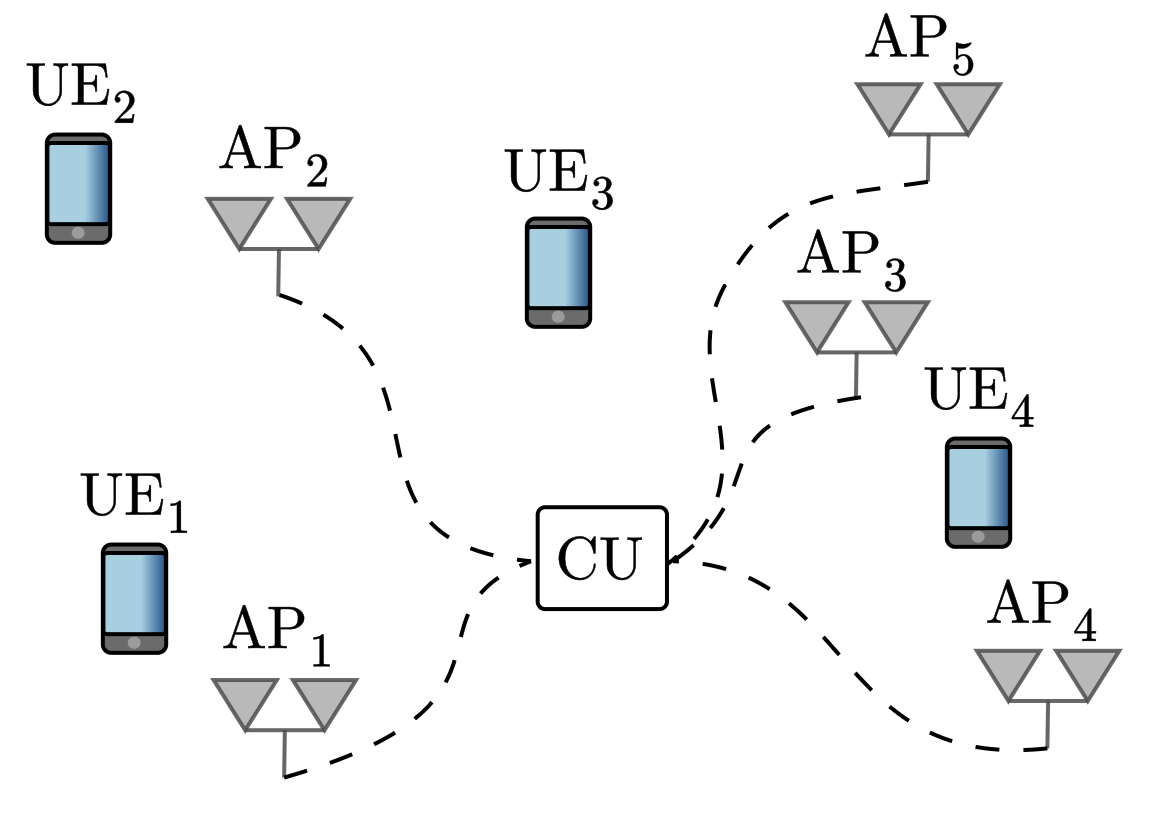}
    \caption{In a cell-free MIMO system, multiple user equipments (UEs) are simultaneously served by a large number of geographically distributed access points (APs). The APs are connected to a centralized unit (CU) via bandwidth-limited fronthaul links, which are used to forward the received signals to the CU for centralized processing.}
    \label{fig:cmimo}
\end{figure}

\label{sec:prob_formulation}

To illustrate the potential benefits of ICL for wireless receivers, we next  consider its application for the task of equalization in cell-free massive MIMO systems.

\subsection{Cell-Free Massive MIMO Equalization}
\label{sec:prob_formulation_cmimo}
As shown in Figure~\ref{fig:cmimo}, we consider the uplink of a cell-free massive MIMO network with $P$ distributed access points (APs), each with $N_r$ antennas, that jointly serve $K$ users, each with $N_t$ antennas. Processing of the received signals is performed at a central unit (CU) that is connected to the APs via fronthaul links with finite capacity \cite{bjornson2019making}.

The pilot sequence received by each AP $m$ is given by
\begin{equation}
    \tilde{\mathbf{y}}_m=\sum^K_{k=1}\mathbf{H}_{m,k}\mathbf{x}_k+\mathbf{n}_m,
    \label{eq:received_ap}
\end{equation}
where $\mathbf{H}_{m,k}\in \mathbb{C}^{N_r\times N_t}$ is the channel matrix from user $k$ to AP $m$ and $\mathbf{n}_m \sim \mathcal{CN}(0,\sigma^2I_{N_r})$ is the additive white complex Gaussian noise. Importantly, due to the simultaneous transmission from the $K$ users, the received signal is a superposition of the transmitted data from all $K$ users, with the further addition of Gaussian noise. We assume that the data symbols of each user $k$ are drawn from a constellation $\mathcal{X}_k$, which may differ across users.

To facilitate the mitigation of inter-user interference, the received signals are forwarded to the CU for joint processing. However, due to the limited capacity of the fronthaul links, each received signal is first quantized using a $b$-bit quantizer $\mathcal{Q}_b(\cdot)$, applied entry-wise and separately to the in-phase and quadrature components \cite{bashar2020uplink}. The CU aggregates the quantized signal ${\mathbf{y}}_m=Q_b(\tilde{\mathbf{y}})$ from each AP $m$, based on which it performs detection of all users’ transmitted symbols $\{\mathbf{x}_k\}^K_{k=1}$.

In this setting, a \emph{task} is defined by the tuple of parameters $\tau=\left(\sigma^2, K,\{\mathcal{X}_k\}^K_{k=1}, \{\{\mathbf{H}_{m,k}\}^K_{k=1}\}^P_{m=1},\right)$, which determines the noise variance $\sigma^2$, the number $K$ of active UEs, the data symbols constellations $\{\mathcal{X}_k\}^K_{k=1}$, and the channels $\{\{\mathbf{H}_{m,k}\}^K_{k=1}\}^P_{m=1}$ .

Equalization for the outlined cell-free massive MIMO receiver faces the following challenges:

\begin{itemize} \item \emph{Limited fronthaul  capacity}: Signal received by the APs are quantized before being sent to the CU, causing non-linear distortions to the signal available at the CU.
\item \emph{Pilot contamination}: In practice, it is typically not possible to assign orthogonal pilots to all active users. The use of non-orthogonal pilots causes pilot contamination, which degrades the capacity of the receiver to separate the contribution of different users \cite{bjornson2017massive}.
\item \emph{Inter-user interference}: Due to the simultaneous transmission from all $K$ users, the received signal is affected by inter-user interference.
\end{itemize}

\subsection{State-of-the-Art}\label{sec:sota}

We now review state-of-the-art methods, starting with conventional two-step equalization.

\subsubsection{Conventional Two-Step Equalization} \label{sec:sota_model_based} Conventional two-step channel equalization methods first estimate the channels $\{\{\mathbf{H}_{m,k}\}^K_{k=1}\}^P_{m=1}$based on pilot symbols forwarded from the APs to the CU, yielding the estimated channels  $\{\{\hat{\mathbf{H}}_{m,k}\}^K_{k=1}\}^P_{m=1}$ and the estimated noise variance $\hat{\sigma}^2$. In the second step, this estimate is used at the CU to equalize the received signal and recover the transmitted data, often through linear processing techniques such as linear minimum mean square error (LMMSE) equalizers, which can potentially account for the presence of quantization noise \cite{bashar2020uplink}.

While this approach is well-understood, versatile, and effective when the channel are well estimated and the underlying linear model is accurate, its performance can degrade significantly in practical scenarios characterized by pilot contamination, short pilot sequences, and non-linear impairments in the radio frequency chain.

\subsubsection{Joint Learning} Machine learning tools can be potentially useful to address the limitations of conventional model-based strategies in the presence of highly non-linear models.  Early attempts to design deep learning-based receivers trained a single model across a range of channels so it could generalize to a variety of channel conditions \cite{samuel2017deep}. 

However, this approach struggles to cover the vast space of channel statistics and front-end configurations, including quantization strategies and fronthaul conditions, and  performance typically degrades for settings that are not well represented in training. 
While joint learning can incorporate channel estimatesthe capacity of adaptation is limited when the estimated channels $\{\{\hat{\mathbf{H}}_{m,k}\}^K_{k=1}\}^P_{m=1}$ are inaccurate, e.g., in the presence of short pilot sequences and front-ends with low-rate quantization. 

\subsubsection{Hypernetworks}

A hypernetwork is a network that generates the weights of another network in response to contextual information. A hypernetwork can be thus trained such that, for given channel estimates $\{\{\hat{\mathbf{H}}_{m,k}\}^K_{k=1}\}^P_{m=1}$, it produces the parameters of an equalization network, thereby sidestepping per-channel training. The equalization network  then processes the received signal $\mathbf{y}$ to estimate the transmitted symbols  $\mathbf{x}$ \cite{raviv2024modular}. The promise of a hypernetwork-based solution is that a single forward pass of the hypernetwork given channel estimates $\{\{\hat{\mathbf{H}}_{m,k}\}^K_{k=1}\}^P_{m=1}$ can  automatically adapt  the weights of the equalizer. 

 As it will be further discussed in Section \ref{sec:icl_results}, hypernetworks can  handle moderate deviations across different tasks without needing retraining \cite{raviv2024modular}. However, their capacity for adaptation is limited by their direct reliance on channel estimates $\{\{\hat{\mathbf{H}}_{m,k}\}^K_{k=1}\}^P_{m=1}$. If these estimates are of insufficient quality, hypernetworks typically struggle to produce well-performing equalization models. Moreover, training the hypernetwork can be complex, as it involves the determination of a high-dimensional output -- the weights of the equalizer -- as a function of a small set of context variables -- the estimated channels.

\subsubsection{ Meta-Learning} Methods like MAML \cite{finn2017model} and their variants \cite{chen2023learning}  train an initial model such that a few gradient steps on a small contextual dataset can yield good performance on a new task. Meta-learning can significantly reduce the training overhead compared to training from scratch for each new channel, particularly when combined with model-based methods \cite{raviv2024adaptive}.  Its main drawback, however, is the reliance of inner-loop local gradient updates. This limits the possible variations across tasks making it hard to accommodate richer task environments. In fact, on the one hand, choosing a smaller learning rate and few updates can facilitate meta-training, but, on the other hand, it hamstrings the adaptation process across different channel conditions.

\subsubsection{Summary} In summary, while joint learning, hypernetworks, and meta-learning each made strides toward the definition of adaptive wireless receivers, they either require reliable channel information  or they can only accommodate limited weight updates  at deployment time. ICL represents a new state-of-the-art methodology that circumvents these limitations by embedding  adaptation within  inference while enabling the direct use of rich contextual information.

\section{ICL-Based Neural Receivers}

\label{sec:icl_receiver}

In this section, we will introduce ICL-based neural receivers for the cell-free massive MIMO system introduced in Section \ref{sec:prob_formulation}, covering also some theoretical properties.

\subsection{ICL-Based Equalization of Cell-Free Massive MIMO Systems}
An ICL-based equalizer estimates the transmitted data symbols by processing, in context, a prompt consisting of the quantized pilot sequences and received data forwarded by the APs to the CU.  In the notation introduced in the previous section, the transmitted symbols corresponds to the target variable $v$, while the received symbols corresponds to the input variable $u$.

\subsubsection{Prompt Design} 
In a cell-free massive MIMO system, each user $k$ is assigned a pilot sequence $\mathbf{p}_k \in \mathbb{C}^{N_t \times T_p}$ of length $T_p$, and we denote $\mathbf{p}_{k,t}$ the $t$-th pilot symbol vector transmitted by user $k$. Since pilot reuse is allowed, the pilot sequences transmitted by the $K$ users are not necessarily orthogonal.

According to the channel model in \eqref{eq:received_ap}, the pilot signals received by each AP $m$, are given by
\begin{equation}
    \tilde{\mathbf{R}}_m = \sum_{k=1}^{K} \mathbf{H}_{m,k} \mathbf{p}_k + \mathbf{N}_m,
    \label{eq:received_pilots_ap}
\end{equation}
where $\mathbf{N}_m \in \mathbb{C}^{N_r \times T_p}$ represents the additive Gaussian noise over the $T_p$ channel uses. Each AP quantizes the received pilot signals using a $b$-bit quantizer and forwards them to the CU. The quantized signals at AP  $m$ are denoted as ${\mathbf{R}}_m = Q_b(\tilde{\mathbf{R}}_m)$. The quantized pilots from all APs are concatenated to form the matrix
\[
{\mathbf{R}} = [{\mathbf{R}}_1^T, \dots, {\mathbf{R}}_P^T]^T \in \mathbb{C}^{(P N_r) \times T_p}.
\]
Each column ${\mathbf{r}}_i \in \mathbb{C}^{(P N_r) \times 1}$ of matrix ${\mathbf{R}}$ corresponds to the received pilot symbols transmitted by the $K$ users at channel use $i$.

Based on the received pilot signals ${\mathbf{R}}$ and knowledge of the pilot sequences, the prompt used to estimate the data symbol $\mathbf{x}_{k}$ for user $k$ is defined as
\begin{equation}
\label{eq:prompt_cmimo}
\{\mathcal{C}_{k}, \mathbf{y}\} = \{ {\mathbf{r}}_1, \mathbf{p}_{k,1}, {\mathbf{r}}_2, \mathbf{p}_{k,2}, \dots, {\mathbf{r}}_{T_p}, \mathbf{p}_{k,T_p}, \mathbf{y} \},
\end{equation}
thus including pairs of received pilot pairs $\{({\mathbf{r}}_i, \mathbf{p}_{k,i})\}^{T_p}_{i=1}$ together with the received signal $\mathbf{y}$.

The prompt in \eqref{eq:prompt_cmimo} can be augmented with additional contextual information, such as indicators of the modulation types used by the users or long-term channel statistics. In Section \ref{sec:exp_pilot_contamination}, we empirically demonstrate that including contextual information -- specifically, long-term scale fading coefficients of users assigned to the same pilot sequence -- can significantly reduce the estimation error in the presence of pilot contamination.
\subsubsection{Architecture}
The prompt in \eqref{eq:prompt_cmimo}, along with additional contextual information, is tokenized and processed by an ICL-capable sequence model, such as a decoder-only Transformer or an SSM (See Section \ref{sec:icl_general}). Tokenization is performed by concatenating the in-phase and quadrature components of the received signals and by converting categorical variables into real-valued quantities using one-hot encoding. The resulting vectors are zero-padded to ensure a uniform length. The $ K $ tokenized sequences -- each corresponding to the prompt \eqref{eq:prompt_cmimo} for user $ k $ -- are then processed in parallel by the Transformer, producing the set of estimated transmitted data symbols $ \{\hat{\mathbf{x}}_{k}\}_{k=1}^K$.
\subsubsection{Meta-Training} Following Section \ref{sec:icl_gen_train}, as illustrated in Figure \ref{fig:icl}, the parameters of the ICL-based equalizer are optimized following these steps:
\begin{itemize}
\item \emph{Task generation:} A set of pre-training tasks $\mathcal{T}_{\rm train}$ is generated by randomly choosing the number and locations of users, and then generating the corresponding channels using standard channel models or ray-tracing simulators. Ideally, the equalization tasks generated during meta-training should be as diverse as possible to encompass configuration similar to those expected at run time.

\item \emph{Prompt generation:} For task $\tau \in \mathcal{T}_{\rm train}$, prompts are obtained by simulating the transmission of pilot sequences and data. The resulting sequences of pilots and received data symbols are used to construct prompt for each user $k$ as in \eqref{eq:prompt_cmimo}, while transmitted data symbols $\{\mathbf{x}_{k}\}^K_{k=1}$ are used as labels to evaluate the equalizer performance.

\item \emph{Optimization:} The loss function compares the estimates $ \{\hat{\mathbf{x}}_{k}\}_{k=1}^K$ with the ground-truth transmitted symbols $\{\mathbf{x}_{k}\}^K_{k=1}$ to compute the mean squared error (MSE) loss 
\begin{align}
\ell(\{\hat{\mathbf{x}}_{k}\}^K_{k=1},\{\mathbf{x}_{k}\}^K_{k=1})=\frac{1}{K}\sum^K_{k=1}\lVert\mathbf{x}_{k}-\hat{\mathbf{x}}_{k}\rVert.
\label{eq:mse_loss}
\end{align}
Alternatively, one can treat also the received pilot $\{\mathbf{r}_i\}^k_{i=1}$ in the prompt \eqref{eq:prompt_cmimo} as a additional query points and using the corresponding transmitted pilots  $\{\mathbf{p}_i\}^k_{i=1}$ as labels to compute the mean squared error loss \eqref{eq:mse_loss} over both reconstructed data and pilot symbols. The equalizer’s parameters are then optimized using standard first-order methods to minimize the estimation errors averaged over the generated tasks and prompts. 
\end{itemize}

\subsection{Theoretical Properties of ICL-Based Equalization}

\label{sec:theoretical_properties}

ICL-based neural receivers can be conceptualized in different ways:  as implicitly carrying out a form of gradient-based optimization \cite{schlag2021linear}, as a form of kernel methods \cite{han2023explaining}, or as  mimicking Bayes-optimal estimators for inverse problems \cite{pmlr-v258-kunde25a}. The latter perspective is particularly relevant for equalization, as the equalizer should ideally produce an estimate of the posterior distribution $p(\mathbf{x}|\mathcal{C},\mathbf{y})$ of the transmitted symbol $\mathbf{x}$ given the corresponding received signal $\mathbf{y}$ and contextual information  $\mathcal{C}$ \cite{pmlr-v258-kunde25a}, or directly evaluate the posterior mean $\mathbb{E}[\mathbf{x}|\mathcal{C},\mathbf{y}]$.  From this viewpoint, an ICL-based equalizer  optimized via meta-training  aims at approximating Bayesian inference. 

Recent work has shown that under suitable conditions (e.g., a linear static channel with a well-defined prior),
in the limit of large context length,
a single-layer self-attention Transformer can provably realize the optimal Bayesian estimator and that the cross-entropy loss is a convex function of the Transformer parameters \cite{pmlr-v258-kunde25a}.
These theoretical results establish that ICL-based equalization can approximate the posterior distribution $p(\mathbf{x}|\mathcal{C},\mathbf{y})$ if it has sufficient capacity and it is given enough training data and pilots. 
These theoretical insights suggest that ICL-based equalization is not a black-box heuristic but a principled approach: it converges to the same solutions as optimal Bayesian inference, achieving the fundamental limits dictated by detection and estimation theory.

\section{Numerical Examples and Discussion}
\label{sec:icl_results}

In this section, we showcase the advantages of ICL-based equalization for a variety of settings drawing from references  \cite{pmlr-v258-kunde25a, zecchin2024cell, zecchin2024context,song2024incontextlearnedequalizationcellfree}.

\subsection{Threshold Behavior of ICL-based Equalization}
\begin{figure}
    \centering
    \includegraphics[width=0.9\linewidth]{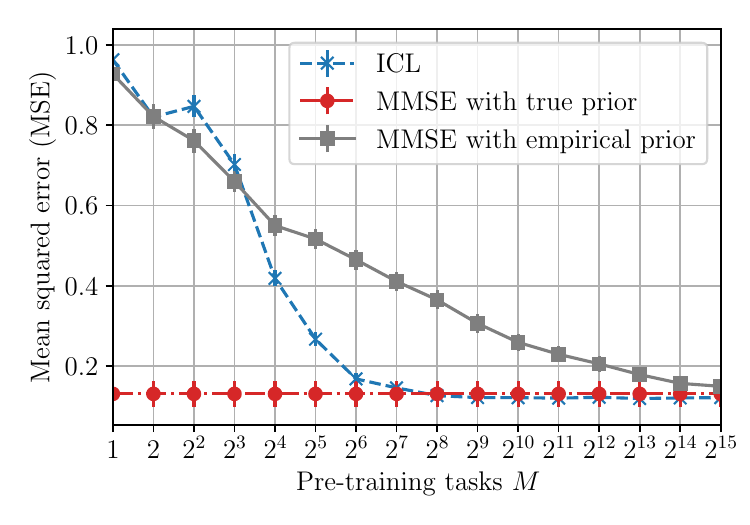}
    \caption{Test MSE as a function of the number of pre-training tasks $M$ of the ICL equalizer and two  MMSE equalizers: one with a uniform prior over the $M$ channel realizations available for training,  i.e., with an empirical prior, and one with a prior that matches the unknown channel distribution $P_\mathbf{H}$.}
    \label{fig:mse_vs_tasks}
\end{figure}

In this section, we study a simplified equalization task to illustrate how ICL-based equalizers behave as approximate Bayesian estimators for the inverse problem of equalization, with priors that depend on the diversity of the meta-training data.  

We consider a basic cell-free MIMO setting, as introduced in Section \ref{sec:prob_formulation}, with a single user and a single AP, both equipped with two antennas. Transmitted symbols are drawn from a unit-power 4-QAM constellation, and the SNR is fixed at 10 dB. The channel matrix  is random $\mathbf{H} $ with i.i.d. complex Gaussian entries. All task parameters are fixed except for the channel matrix $\mathbf{H}$, which varies independently across tasks. 

Figure \ref{fig:mse_vs_tasks} reports the test MSE  of an ICL equalizer implemented via a Transformer with $L=2$ layers, $H=4$ attention heads, and embedding dimension $D_e=64$, which is trained on $M$ realizations of the channel $\mathbf{H} \sim P_{\mathbf{H}}$. For comparison, we also plot the MSE attained by two MMSE equalizers, one with an empirical prior evaluated using the $M$ channel realizations available for training, and one with the true prior $P_{\mathbf{H}}$. The performance is seen to improve with an increasing number of equalization tasks $M$, as the model is exposed to a broader range of channel conditions. Furthermore, the MSE achieved by ICL exhibits a threshold behavior: for small number of tasks $M$, the performance follows  that of the MMSE estimator with the empirical  prior, while, for large $M$, the MSE approaches that of the MMSE solution with a prior matching the true channel distribution $P_\mathbf{H}$. This demonstrates that the ICL framework can extrapolate beyond the training distribution and recover near-optimal Bayesian behavior when trained on a sufficiently diverse set of tasks. This results aligns with the theoretical properties of ICL-based equalization discussed in Section \ref{sec:theoretical_properties}.

\subsection{Comparison with Alternative Frameworks}
\label{sec:exp_icl_vs_sota}
\begin{figure}
    \centering
    \includegraphics[width=0.9\linewidth]{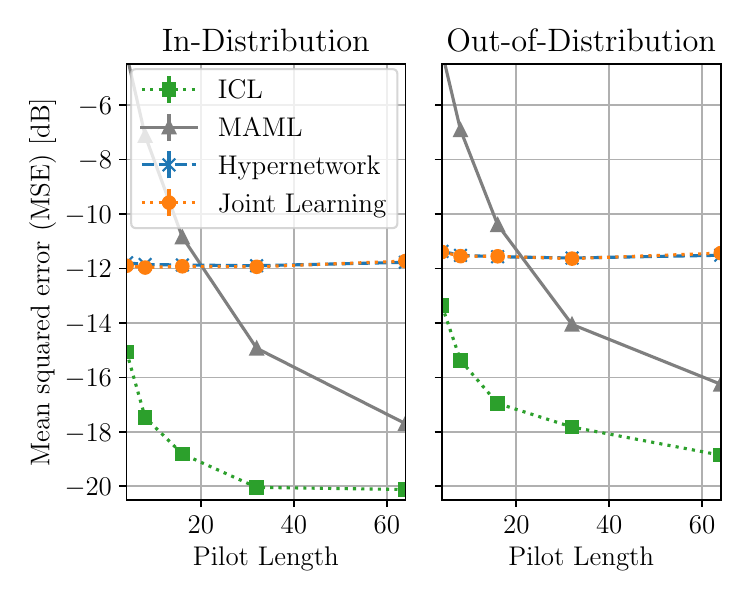}
    \caption{Test MSE as a function of pilot symbol sequence length of neural receivers based on joint learning, meta-learning, hypernetworks, and ICL. The left plot shows results on in-distribution tasks, while the right plot shows results on out-of-distribution tasks, obtained by increasing the noise power by 10 dB.}
    \label{fig:sota_comparison}
\end{figure}
We now consider a more complex cell-free network and compare the performance of the ICL-based receiver against existing neural receivers based on joint learning, hyper-networks, and meta-learning.

Following a setup similar to \cite{bjornson2019making}, we simulate a cell-free massive MIMO network of $1 $ km $\times 1$ km square area with $P=4$ APs, each equipped with $N=2$ antennas. Communication occurs at a 2 GHz carrier frequency, with path loss and spatially correlated fading modeled using the 3GPP Urban Microcell standard \cite{3gpp} and the Gaussian local scattering mode \cite{bjornson2019making}. We assume a random number of active users $K$, uniformly distributed between 1 and, which are  placed uniformly at random across the deployment area. Users are assigned pilot sequences, which for the moment are assumed to be orthogonal, and they generate data symbols from BPSK, 8-PSK, 4-QAM, 16-QAM and 64-QAM constellations. The noise power is fixed to $\sigma^2 = -146$ dB, yielding an average SNR of 24 dB across the deployment area. We consider a fixed fronthaul capacity of $b=4$ bits per symbol.

 We consider a meta-training dataset comprising $N_{\rm tr} = 8192$ possible network deployments, and for each deployment we generate $N_{\rm ex} = 1024$ prompts by simulating the channel propagation and transmission of pilot sequences. For this dataset, we train an ICL-receiver based on the Transformer model with $L=4$ layers with $H=4$ attention heads and embedding dimension $D_e=64$, and compare it against neural receivers based on joint-learning, model-agnostic meta-learning (MAML) \cite{finn2017model} and hypernetworks \cite{raviv2024modular}. The model sizes are chosen such that their inference-time computational complexity, measured in FLOPs, is on the order of $10^5$ to $10^6$, making them comparable to the ICL equalizer.
 
 In Figure~\ref{fig:sota_comparison}, we report the test MSE as a function of the pilot sequence length when tested on data from the same distribution used for training, as well as on out-of-distribution data obtained by increasing the noise power by $10$ dB. In both scenarios, as the pilot sequence length increases, more information about the channel becomes available for adaptation. The ICL-based receiver effectively leverages this information, reducing its MSE as the sequence grows and outperforming all other baselines. The MAML-based receiver, which uses first-order methods to adapt to the current channel, performs comparably to the ICL-based receiver only when the pilot sequence is sufficiently long. In contrast, neural receivers based on hypernetworks and joint learning either fail to converge or lack the adaptability needed to improve performance when provided with additional task information. 
 
 When tested on out-of-distribution data, the performance of all methods decreases due to the mismatch between training and testing conditions and the increased noise power. However, even for out-of-distribution testing, the ICL-based receiver outperforms all other neural baselines. These findings underscore the superior generalization power and adaptability of ICL-based receivers compared to existing neural approaches, especially for short pilot sequences.
 
\subsection{Transformers vs. SSMs}

In the following, we adopt the simulation setup from the previous section to compare the performance, memory, and computational requirements of ICL-equalizers implemented using Transformers, referred to as T-ICL, and SSMs, i.e., SSM-ICL. As mentioned in Section \ref{sec:icl_general}, Transformer-based ICL relies on the parallel processing of the received prompt, whereas SSMs are inherently sequential. This results in distinct computational complexities at test time: quadratic in the input sequence length for Transformers and linear for SSMs.

\begin{figure}
    \centering
    \includegraphics[width=0.9\linewidth]{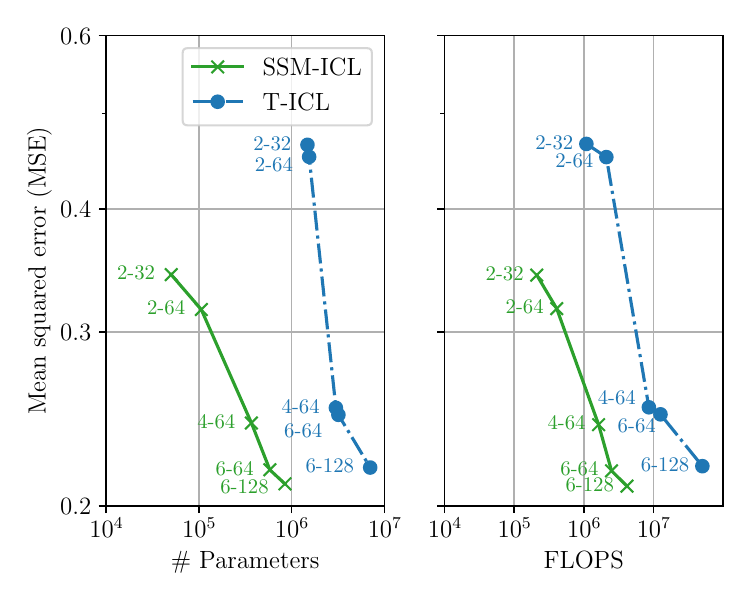}
    \caption{Test MSE obtained by ICL equalizers implemented using Transformers and SSMs as a function of number of parameters and FLOPS. The curves are obtained by  varying the numbers of layers and hidden dimensions in the feedforward neural networks. For each model, the first number denotes the number of layers, while the second denotes the hidden dimension.}
    \label{fig:params}
\end{figure}

For Figure \ref{fig:params}, we train various Transformer and SSM architectures by varying the number of layers and the hidden dimension of the feedforward neural networks used in the embedding operations. For each architecture, we evaluate the number of model parameters, which dictates the memory requirements, as well as the number of floating-point operations (FLOPS) required at test time, a measure of computational complexity. We specifically plot the MSE against number of parameters and the  number of FLOPS under the same testing conditions as in the previous section. 

In general, larger network architectures -- whether Transformer- or SSM-based -- lead to reduced MSE due to their increased capacity. However, for the same MSE level, SSMs are shown to significantly reduce both the number of parameters and the FLOPs required for inference. This highlights the superior scalability of SSMs compared to Transformers, emphasizing their suitability for deployment on resource-constrained devices.

\subsection{Impact of Fronthaul Capacity and Pilot Contamination}

In Section \ref{sec:exp_icl_vs_sota}, ICL-based receivers were shown to outperform both model-based and data-driven approaches in the case of short pilot sequences, while assuming orthogonal pilot assignments.  In the following, we investigate the performance of the ICL equalizer under strong nonlinear impairments and pilot contamination. The aim is to demonstrate how incorporating contextual information into the prompt can enhance equalization performance under these challenging conditions.

\begin{figure}
    \centering
    \includegraphics[width=\linewidth]{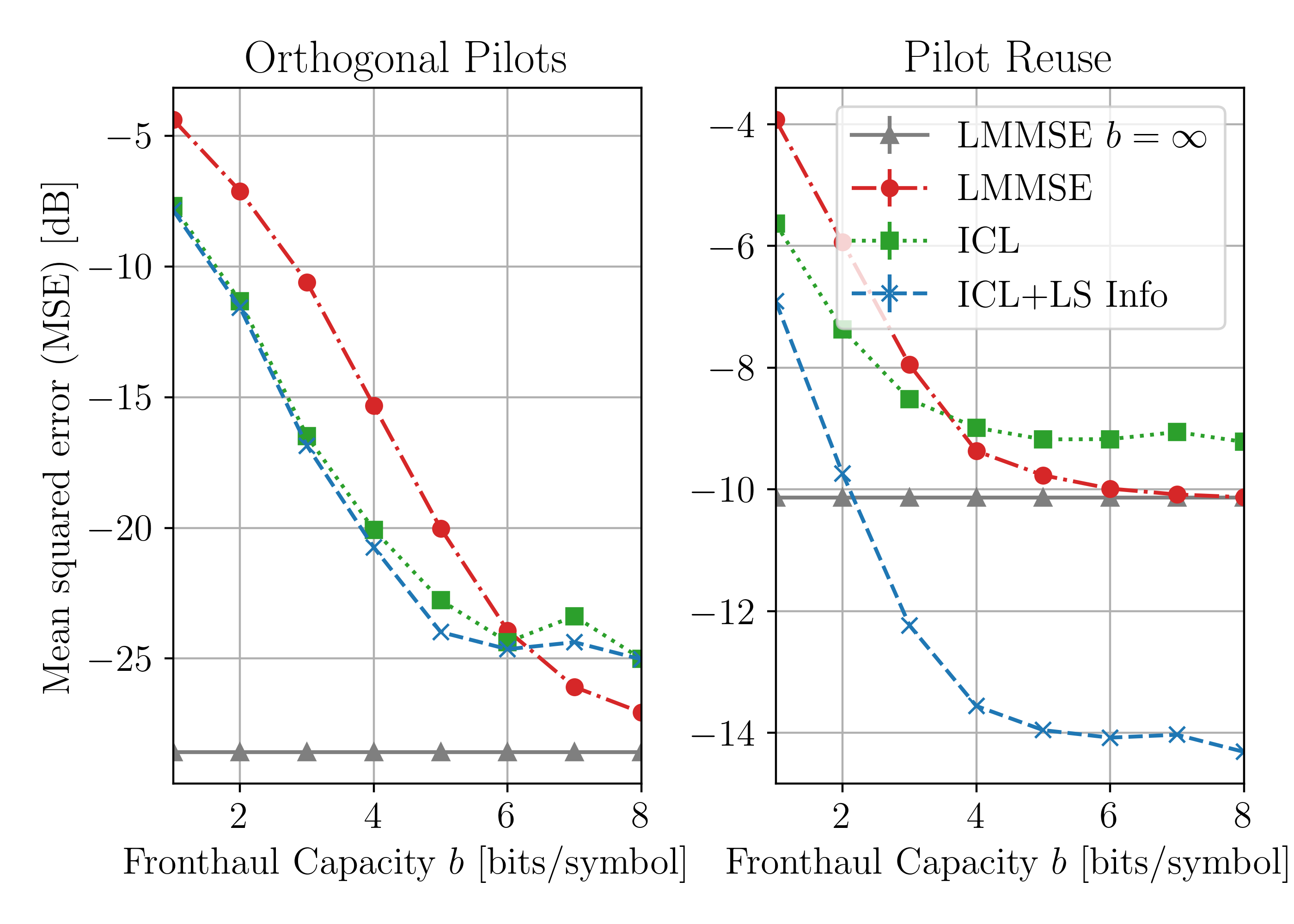}
    \caption{Test MSE with varying fronthaul capacity under orthogonal pilot assignment (left) and pilot reuse (right). We compare the centralized LMMSE solution -- with both limited and unlimited fronthaul capacity -- against two ICL equalizers based on Transformers: one using only in-context examples and another augmented with large-scale fading tokens. Both ICL-equalizers operate under a finite fronthaul capacity.}
    \label{fig:pilot_reuse}
\end{figure}

Specifically, we consider a cell-free MIMO system with varying fronthaul capacity $b$ and pilot reuse. Under pilot reuse, pilots are assigned uniformly, potentially leading to the same pilot being assigned to multiple users. Figure \ref{fig:pilot_reuse} reports the MSE of an ICL-based equalizer in comparison to a centralized linear MMSE equalizer \cite{bashar2020uplink}. We evaluate two versions of the ICL equalizer: one in which the prompt includes only in-context examples, and another where the prompt is augmented with long-term channel fading coefficients of users sharing the same pilot sequence. Note that these coefficients are also required for implementing the linear MMSE equalizer, ensuring that both methods are provided with equivalent information.

As shown in Figure \ref{fig:pilot_reuse}, in the case of orthogonal pilots, both ICL equalizers perform in a similar way,  outperforming the LMMSE equalizer when fronthaul capacity is limited. As the fronthaul capacity $b$ increases, all methods converge toward the performance of the ideal LMMSE solution with unquantized observations (i.e., $b = \infty$). 

However, in the presence of pilot contamination, the ICL equalizer without access to long-term fading coefficients is unable to mitigate the effects of pilot contamination and performs comparably to the LMMSE equalizer. In contrast, the ICL equalizer augmented with long-term fading information significantly reduces the MSE across all fronthaul capacity levels, even outperforming the LMMSE solution with perfect fronthaul. This improvement is due to the fact that long-term channel statistics allow the ICL equalizer to resolve the ambiguity caused by pilot reuse and effectively separate their contributions in the received signal. This highlights the capacity of ICL to exploit contextual information -- here, long-term fading coefficients -- to improve equalization performance and surpass conventional model-based approaches.

\label{sec:exp_pilot_contamination}

\section{Outlook}\label{sec:outlook}

The application of ICL to receiver design is still in its early stages. Looking ahead, several exciting directions can be foreseen to enhance and expand the capabilities of ICL-based receivers.

\subsubsection{Hardware-Efficient Implementations} The quadratic complexity of Transformers in the context size may be a bottleneck for the implementation of ICL, particularly on mobile devices. Techniques like quantization can  reduce model size to mitigate this problem \cite{liu2021post}. As a complementary direction,  \emph{neuromorphic computing}, involving spiking neural networks (SNNs) and event-driven processing, can be leveraged for efficient implementations of Transformers \cite{song2024neuromorphic}.

\subsubsection{Integration with Channel Coding} In a real communication chain, symbol detection is followed by channel decoding corresponding to a given error-correcting code. There is potential to integrate ICL-based equalization with the decoding process via a ``turbo'' ICL process, which uses decoded bits from a channel code to refine contextual information \cite{song2025turbo}. 

\subsubsection{Adaptive Meta-Training} Future wireless systems may face scenarios in which the environment changes unexpectedly owing to the integration of technologies such as intelligent reflecting surfaces. In these cases, it would be useful to have ways to recognize that the current tasks are outside the distribution of tasks encountered during meta-training, calling for further model optimization. This approach could lead to an adaptive meta-learning procedure that opportunistically adapts the weights of the ICL model.

\subsubsection{Beyond Equalization} ICL could be extended to tasks beyond equalization. Examples may include  beam selection, where a context of  received power measurements may let a model pick the best beam, or scheduling, in which contextual information about  traffic patterns may be mapped to a scheduling policy. 

\subsubsection{Theoretical Foundations} While, as discussed, there has been some work on the theoretical analysis of ICL in wireless systems, there is a need for more research on this topic.  Open questions include the study of inverse problems involving non-linear observations models.

\subsubsection{Security} If a malicious agent manipulates the context, e.g., via pilot contamination attacks, how would an ICL model respond? Could it be more vulnerable than a conventional method, or could it rather be more robust because it does not rely on explicit channel estimates?

\subsubsection{Integration with Intent-based Networking} Sequence models such as Transformers and SSMs can be trained to efficiently process text. This suggests that the use of ICL in wireless systems could be integrated with \textit{intent-based} networking, in which natural language requirements are interpreted by the ICL model as part of its prompt.

\section{Conclusions}
\label{sec:conclusions}
ICL offers a powerful paradigm  for the design of adaptive wireless receivers. By leveraging large sequence models to interpret pilot and context information on the fly, ICL-based receivers can achieve gradient-free adaptation to changing channels and interference conditions, all within a single forward pass. In this paper, we have outlined the application of this approach to cell-free massive MIMO systems, showing that ICL not only bridges the gap between meta-learning and hypernetwork strategies but it also outperforms classical two-step solutions in challenging scenarios. 

The key principle underlying ICL is that a sequence model can ``learn to learn'' from examples without explicit fine-tuning. From a theoretical perspective, the capacity of Transformers to implement learning algorithms internally provides a basis for understanding ICL’s success. Practically, the examples reviewed demonstrate substantial gains, confirming the significance of ICL for next-generation communication systems. 

As models and training techniques evolve, and as we gain more theoretical insights, we anticipate that ICL will become a mainstay in the toolbox of wireless communication engineering, driving the development of robust and self-adapting networks in next-generation wireless systems.

\bibliography{IEEEabrv,biblio}
\bibliographystyle{ieeetr}
\vspace{-1em}

\end{document}